\documentclass[10pt,journal,twoside]{IEEEtran}
\usepackage{longtable}
\usepackage{amsmath,epsfig,times}
\usepackage[T1]{fontenc}
\usepackage{latexsym,epsf,graphicx,psfrag,amssymb,cite}
\usepackage{cite,tabularx}
\usepackage{epsf}
\usepackage{amsmath,tikz}
\usepackage{amsmath}
\usepackage{stfloats}
\usepackage{amssymb}
\usepackage{epsfig}
\usepackage{psfrag}
\usepackage{tikz}

\usepackage{graphicx}
\usepackage{multirow}
\usepackage{array}
\usepackage{enumerate}
\usepackage{algorithm}
\usepackage{algorithmic}
\usepackage{amsthm}
\usepackage{float,lscape}
\usepackage{longtable}
\usepackage{caption}
\usepackage{url}
\usepackage{makecell}
\usepackage{soul}
\usepackage[caption=false]{subfig}

\begin{document}
\title{UAV-Empowered Disaster-Resilient Edge Architecture for Delay-Sensitive Communication}
\author{Zeeshan~Kaleem, Muhammad Yousaf, Aamir Qamar, Ayaz Ahmad~\IEEEmembership{Senior Member,~IEEE}, Trung Q. Duong~\IEEEmembership{Senior Member,~IEEE}, Wan Choi~\IEEEmembership{Senior Member,~IEEE}, Abbas Jamalipour~\IEEEmembership{Fellow,~IEEE}
\thanks{Zeeshan~Kaleem, Aamir Qamar, Ayaz Ahmad are with the Department of Electrical and Computer Engineering, COMSATS University Islamabad, Wah Campus, Pakistan.    (e-mail: zeeshankaleem@gmail.com, aamirqamar@ciitwah.edu.pk, ayaz.uet@gmail.com)

Muhammad Yousaf is with the Department of Telecommunications Engineering, UET Taxila, Pakistan. e-mail: myousafalamgir@yahoo.com 

Trung Q. Duong is with the School of Electronics, Electrical Engineering and Computer Science, Queen's University Belfast, UK. (email: trung.q.duong@qub.ac.uk)

Wan Choi is with the School of Electrical Engineering, KAIST, Korea. (email: wchoi@kaist.edu)

Abbas Jamalipour is with the School of Electrical and Information Engineering, University of Sydney, Australia. (email: a.jamalipour@ieee.org)
}}
\maketitle
\begin{abstract}
The fifth-generation (5G) communication systems will enable enhanced mobile broadband, ultra-reliable low-latency, and massive connectivity services. The broadband and low-latency services are indispensable to public safety (PS) communication during natural or man-made disasters. Recently, the third generation partnership project long term evolution (3GPP-LTE) has emerged as a promising candidate to enable broadband PS communications. In this article, first we present six major PS-LTE enabling services and the current status of PS-LTE in 3GPP releases. Then, we discuss the spectrum bands allocated for PS-LTE in major countries by international telecommunication union (ITU). Finally, we propose a disaster resilient three-layered architecture for PS-LTE (DR-PSLTE). This architecture consists of a software-defined network (SDN) layer to provide centralized control, an unmanned air vehicle (UAV) cloudlet layer to facilitate edge computing or to enable emergency communication link, and a radio access layer. The proposed architecture is flexible and combines the benefits of SDNs and edge computing to efficiently meet the delay requirements of various PS-LTE services. Numerical results verified that under the proposed DR-PSLTE architecture, delay is reduced by 20\% as compared with the conventional centralized computing architecture.
\end{abstract}
\section{Introduction}\label{Introduction}
The fifth-generation (5G) communications system is targetting broadband and ultra-reliable communications to meet users' diverse requirements. During a disaster, public safety (PS) organizations such as police (road, railway, and airport), military, guards (border, coastal, customs), and hospitals demand broadband and low-latency communications to timely provide emergency services. Currently, PS organizations rely on narrow band Terrestrial Trunked Radio (TETRA)-based systems which can only support voice services. By enabling fast and broadband communications among PS users and organizations, real-time video can be transmitted to increase the recovery chances of PS users \cite{3baldini_survey_2014}. To achieve high efficiency and replace the existing TETRA networks, third generation partnership project long-term evolution (3GPP-LTE) broadband standard is adopted to support voice and broadband video applications to meet the PS users' demands. Moreover, to enable PS communication, unmanned air vehicle (UAV) have received attention from academia and industry. UAVs are the best possible enabler for PS communications as compared with existing static terrestrial networks because they can be easily integrated anywhere in the disaster-hit areas. For instance, in \cite{mekikis2017communication} authors proved the benefits of integrating UAVs during disaster. They showed that how the altitude and distance between UAV effects the network recovery probability, and also validate the proposal efficacy by simulations and experimentally.

The 3GPP objective is to upgrade the existing LTE-based architecture in order to enable broadband PS communications (PSC). To upgrade the existing services, 3GPP TS 22.179 Release 13 \cite{access20093gpp} has recently introduced mission-critical push-to-talk (MCPTT) services. In this release, in addition to enabling direct mode communication, the MCPTT also adds the feature of neighbor discovery which may be network assisted or without network assistance. Moreover, relay capability has also been included to provide service to out-of-coverage users. To achieve these targets, we need a promising architecture that can meet the desired bandwidth and latency requirements during emergency situations. Hence, to overcome these limitations, we have three major contributions: 1) we presented six major PS-LTE enabling services and the current status of PS-LTE in 3GPP releases, 2) discuss the spectrum bands allocated for PS-LTE in major countries by international telecommunication union (ITU), and 3) finally we propose a disaster resilient three-layered architecture for PS-LTE (DR-PSLTE), namely: software-defined network layer, UAV cloudlet layer \cite{jeong2017mobile}, and radio access network layer.

\section{Evolution of TETRA-based Public-Safety Network and Related Works}\label{RW}
Recently, there has been growing interest in improving communication technologies for PS networks and the existing TETRA-based systems. The major motivation is to provide PSC services till the completion of the evolution phase from the TETRA to the LTE. The evolution plan in Europe, specifically Finland, has five main steps to completely implement hybrid networks that could achieve almost similar performance like broadband LTE for emergency communication services. These steps are: 1) a mobile virtual network operator (MNVO) set up to meet increased data rate requirements. For this purpose, initially, the externally available broadband services will be used for PS services which will be replaced by the LTE core in the future, 2) critical content will be served via TETRA, while non-critical content will be offered via broadband communications, 3) dedicated broadband in some areas will start functioning for PS services using LTE core, 4) excellent voice services will be provided in TETRA and LTE-based networks, 5) LTE-based broadband services will completely take over the TETRA-based services after the completion of evolution from TETRA to LTE.

To achieve the aforementioned goals, future PS standards needs to be developed based on LTE. For example, in \cite{doumi2006spectrum}, the major driving factors for wideband spectrum allocation for public safety agencies, and the pros and cons of each available frequency bands are summarized.  Moreover, in \cite{kaleem2016public} authors proposed device-to-device (D2D) discovery scheme for proximity services which has capability of discovering around 63\% more users as compared with random access schemes with the same allocated bandwidth. 
Furthermore, major enabling communication technologies for PS-LTE were discussed in \cite{2kumbhar_survey_2017}. This discussion concludes that the improvement in spectral efficiency of PS networks can be potentially achieved by introducing millimeter wave (mmWave) band, massive multiple-input and multiple-output (MIMO), and UAVs. can be the potential candidates for PSC.  However, they overlooked the detailed discussions on the current status of 3GPP PS-LTE, and a suitable architecture for PS-LTE systems.

\begin{table*}[!hb]
\centering
\caption{Spectrum Allocation for PSC.}
\label{Spectrum For PS}
\begin{tabular}{|p{0.563in}|p{1.063in}|p{1.567in}|l|}\hline
\textbf{ITU Regions}&\textbf{Major Countries}&\textbf{Frequency Band}& \textbf{Bandwidth (MHz)}\\ \hline
Region 1& Europe & 410-430/450-470 MHz (400 MHz) & 40 (20MHz + 20MHz)\\ 
~&~& 733/758-788 MHz (700MHz)& 60 (30MHz + 30MHz)\\
~& UK &No dedicated band (Uses commercial LTE bands)&  \\ \hline
Region 2& Americas &\text{25-50 MHz} \text{(VHF Lower Band)}&6.3 MHz\\ 
~&~& 150-174 MHz (VHF Upper Band)&3.6 MHz (non-contiguous)\\ 
~&~&220-222 (220 MHz band)&0.1 MHz\\ 
~&~&450-470 (UHF Band)&3.7 MHz (non-contiguous)\\ 
~&~&470-512 MHz (T-Band)&6 to 12 MHz blocks (contiguous in specified markets)\\ 
~&~&758-769/788-799 MHz&22 MHz(11 MHz + 11 MHz)(contiguous)\\ 
~&~&768-775/798-805 (700 MHz)&14 MHz (7 MHz + 7 MHz) (contiguous)\\ 
~&~&\text{806-809/851-854 MHz}&6 MHz(3 MHz + 3 MHz) (contiguous)\\ 
~&~&\text{809-815/854-860 MHz (800 MHz)}&3.5\text{ MHz} (\text{1.75 MHz + 1.75 MHz}) (non-contiguous)\\ 
~&~&4940-4990 MHz (4.9 GHz)&50 MHz (contiguous)\\ 
~&~&5850-5925 MHz band (5.9 GHz)&75 MHz (contiguous)\\ \hline
Region 3& Australia &4940-4990 MHz (4.9 GHz)&50 MHz (contiguous)\\ 
~& Japan&4940-4990 MHz (4.9 GHz)&50 MHz (contiguous)\\
~& South Korea & 718-728/773-783 (700 MHz)&20 MHz (10 MHz + 10 MHz)   \\ \hline
\end{tabular}
\end{table*}

In \cite{carla2016lte}, the authors extended the existing LTE-based architecture for the provisioning of multimedia services to the life saving officers. The main challenge was to exploit the benefits of broadband services by satisfying PS service requirements. To enable multimedia services, major steps involved in architecture evolution from the shared commercial and PS network to the independent PS network are also presented. They activated dynamic evolved multimedia broadcast multicast service (eMBMS) to improve the spectral efficiency. Simulations results proved that the proposed system's spectral efficiency is high. However, this architecture could not meet the strict low-latency requirements and is not flexible enough to be centrally updated. To solve this challenge, we propose a three-layered DR-PSLTE architecture that can meet the strict latency requirements by processing important functions at the edge and can also be centrally managed using SDN functionality.

In this context, this article provides a complete overview of PS-LTE and design guidelines on deploying PS-LTE systems. In particular, our major contributions are threefold; 1) we review the current status of PS-LTE in the 3GPP standard releases, 2) we present the communication scenarios and the key enablers for PS-LTE, 3) we propose a DR-PSLTE architecture that suits well for the emergency situations.

\section{PS-LTE Spectrum Allocation and 3GPP Releases Status}\label{ES}

In this section, we summarize the spectrum bands allocated for PSC in major countries located in three different regions, and the status of the 3GPP releases related to PS-LTE.

\subsection{PS communications spectrum allocation}
In 	PSC, delay and bandwidth requirements of emergency services differ from each other. The sharing of the existing available limited spectrum among PS users and non-PS users is not an efficient and sufficient solution to meet the demands of emergency services. Therefore, a dedicated broadband spectrum is desired for PS users. ITU divides the world into three ITU regions to efficiently manage the spectrum. The frequency bands reserved for PSC in the major countries are summarized in Table \ref{Spectrum For PS}. The band allocation has two main categories: contiguous and non-contiguous, in which adjacent and non-adjacnet frequency bands are allocated, respectively. The world radio conference (WRC-15) with Resolution 646 is the agreement between united nations and ITU \cite{2kumbhar_survey_2017} in which they encouraged the PS organizations to use frequency range 694-894 MHz for broadband PSC. Since different countries have different operational frequency ranges and spectrum requirements, the dedicated bands in very high frequency (VHF), ultra high frequency (UHF), 700 MHz, 800 MHz, and 4.9 GHz bands might be used to enable PSC in most of the countries. For spectrum harmonization across the world, dedicated broadband spectrum for PSC is desired.

South Korean government plans to build a dedicated broadband network for PSC. They plan to reserve 20 MHz dedicated spectrum in the 700 MHz band. Similarly, other countries in Region 3 such as Japan and Australia also reserved dedicated spectrum for this purpose. On the other hand, UK did not reserve any specific bands for PSC, but has decided to use the existing LTE bands.

\subsection{Status of PS-LTE key enabling services in 3GPP releases}
The 3GPP has given high priority to PSC as it is rapidly evolving by taking requirements input from the global critical communications industry.  In the following, we provide an overview of the developments achieved for PSC in 3GPP, and also provide a brief overview of the goals to be achieved in the future. 

In 3GPP Release 12 key enablers such as ProSe, group communication, mission critical services, and public warning systems (PWS) for PSC are introduced. In Release 13, the first technical specifications (TS) of Isolated E-UTRAN operation for public safety (IOPS) are determined, whereas more MC communication services like MC data and video are introduced in Releases 14 and 15, respectively. These services are helpful to cope with the disasters by providing emergency communication to the PS officers. 

\begin{figure}[!ht]
 \centering
  \begin{center}
    \includegraphics[width=9cm,height=9.5cm,angle=0]{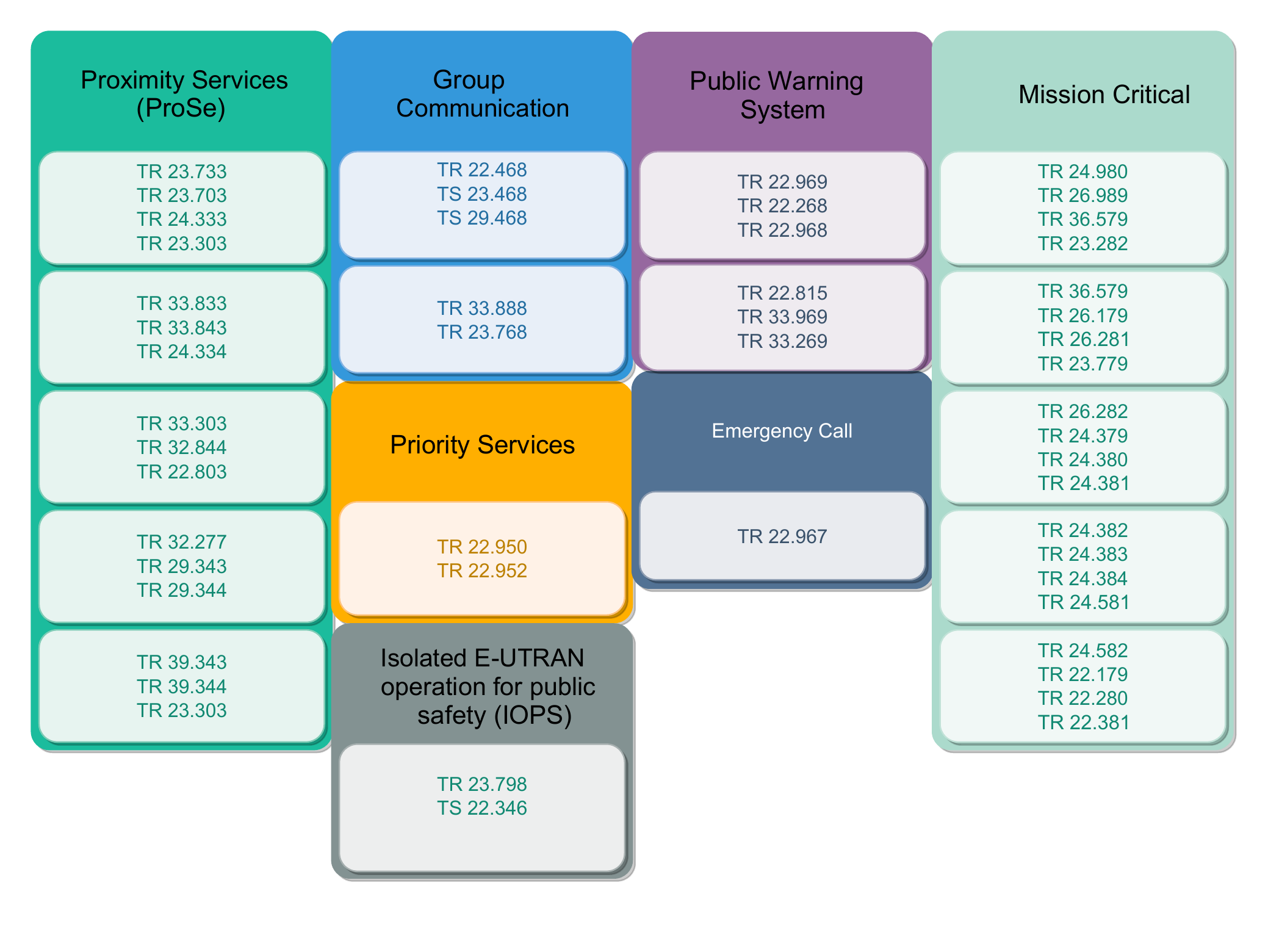}
  \end{center}
  \caption{3GPP technical specification and reports on PS services.}
  \label{PS-ltespecs}
\end{figure}

Since Release 11, 3GPP has led the development of TSs in cooperation with PS industry partners to adapt LTE according to PS requirements. Several TSs have been released for different services. For example, ProSe and group communication system enabler (GCSE) are introduced in 3GPP TS 22.278 and 3GPP TS 22.468 of Release 12, respectively. Similarly, in Release 13, MCPTT and IOPS are adopted and explained in 3GPP TS 22.179 and 3GPP TS 22.346, respectively. In Fig. \ref{PS-ltespecs}, we briefly summarize most of the PS-LTE related 3GPP TS and technical report (TR) documents for ProSe, GCSE, PWS, and MC services which would be helpful for researchers starting their research in PS-LTE.

\section{Public Safety Commmunication Service Key Enablers}\label{ET}
In this section, we discuss the PSC service key enablers such as ProSe, GCSE, MCPTT, IOPS, and the priority services. Moreover, we also describe their needs in PS-LTE system with key components necessary for their implementation.
\subsection{ProSe in PS-LTE}\label{Proximity Services(ProSe)}
3GPP supports ProSe in PS-LTE to enable direct communications among neighboring users as shown in Fig. \ref{PS Architecture}(a). 3GPP ProSe features consist of ProSe discovery and ProSe direct communication. The former finds the neighboring users by broadcasting a beacon whereas the ProSe direct communication enables to establish communication links among the discovered users that lie in the surroundings\cite{access3gpp}. 
To enable ProSe discovery in LTE-based systems, users require 360 KHz of bandwidth to transmit the information. In ProSe group and broadcast communications, data is transmitted by using one-to-many communication approach and that feature can work without discovering the nearby users. ProSe relay is beneficial in users' range extension and there can be either user-to-user relay or user-to-network relay.

\subsection{Group communication system enabler for PS-LTE }\label{Group Communication}
\begin{figure}[!t]
 \centering
  \begin{center}
    \includegraphics[width=9cm,height=9cm,angle=0]{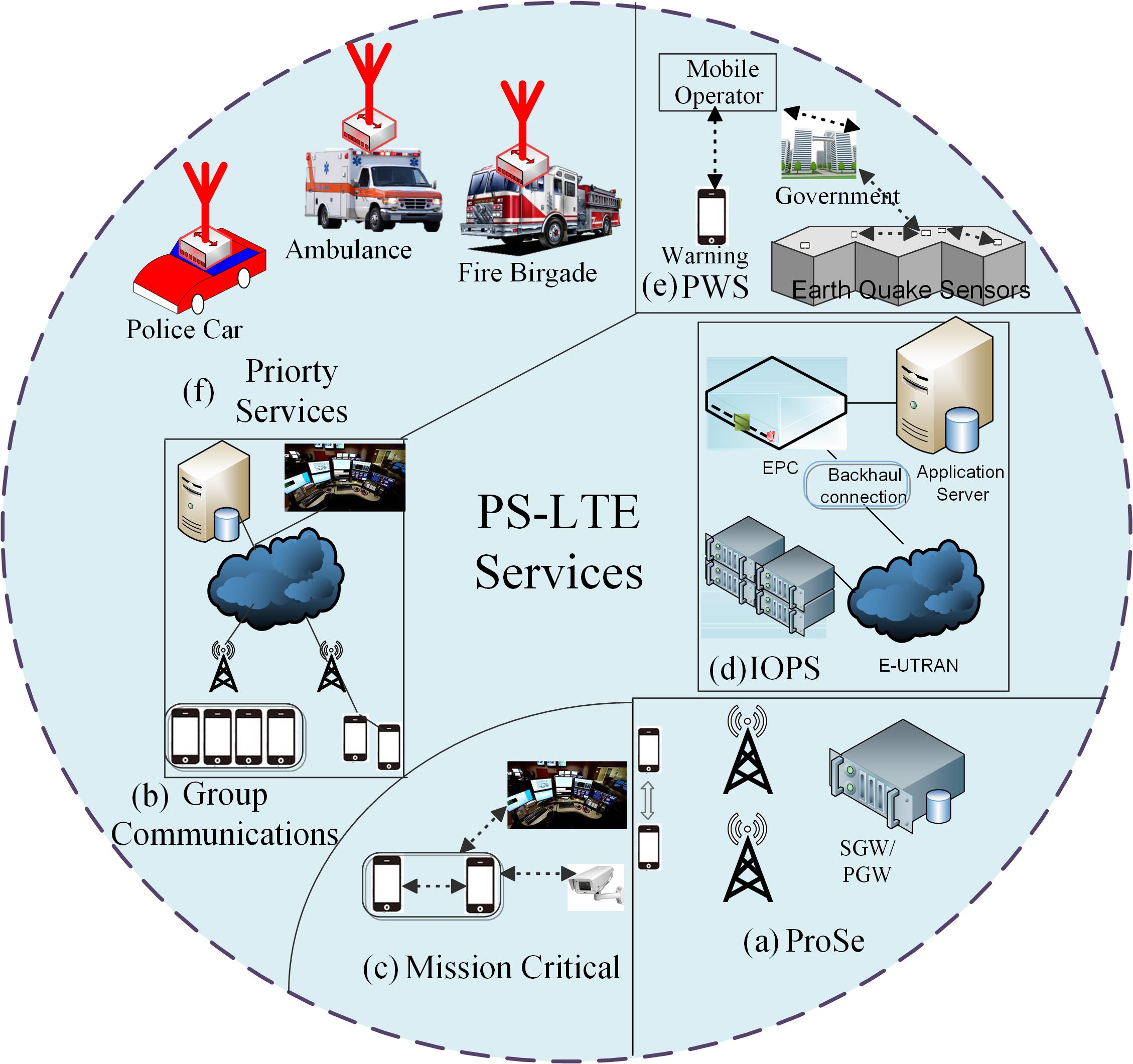}
  \end{center}
  \caption{PS-LTE services: (a) proximity services (ProSe); (b) group communication system enabler (GCSE); (c) mission critical communication; (d) Isolated evolved UMTS terrestrial radio access network (E-UTRAN) operation for PS (IOPS); (e) public-warning system (PWS); (f) priority services.}
  \label{PS Architecture}
\end{figure}

LTE has the capability to provide users with very high data rates. High data rate provisioning can also be helpful for the PS offices and users willing to share multimedia contents. To make use of such a high speed data, 3GPP in Release 12 introduced GCSE by upgrading the existing LTE architecture. The GCSE implementation requires a group call enabler for LTE network, enhanced nodeBs (eNBs), user equipment (UE) relays, group members, group call application server, and dispatcher as depicted in Fig. \ref{PS Architecture}(b). The GCSE supports a large number of group members, enables the push to talk, and offers low latency in communication. GCSE is an efficient approach to convey the same information to multiple users. This system is already functional in the conventional land mobile radio system (LMR) with the feature named as push to talk (PTT). In the conventional system, PTT has only the feature of transmitting the voice to a single group of users. In LTE-based networks, GCSE is expected to support video, voice, and even data communications. In LTE, GCSE users can communicate to several groups in parallel by transmitting voice to one group and video to another. In GCSE system to ensure a group call quality, the group call setup time and the group data dissemination should take less than 300 ms including the request generation time and the receiving time of the packet. In a bandwidth perspective, channel bandwidth would be 180 KHz due to LTE architecture whereas exact bandwidth varies from user to user.

The GCSE architecture is developed on top of the existing LTE architecture. The GCSE architecture is broadly categorized into two layers: a 3GPP evolved-packet system (EPS) layer and an application layer. The EPS layer has main components: mobility management entity (MME), policy and charging control (PCC) function, serving/packet data network (PDN) gateway (S/P-GW), and home subscriber server (HSS). In addition to these components, the GCSE enabled LTE architecture includes multimedia broadcast multicast service (MBMS). Details of each components are provided in Section \ref{architecture}. The application layer runs on top of the 3GPP EPS layer which contains the client applications running in mobile users and the application server on the network side.

\subsection{MCPTT in PS-LTE}\label{Mission Critical}
Fig. \ref{PS Architecture}(c) shows the mission critical communication services that provides emergency communication services during a disaster in lieu of malfunction base station. D2D communication links will be established to provide alternative communication links. For instance, wireless video surveillance camera can send data towards PS officers without using a base station by enabling D2D communications. The main target of MCPTT is to enable reliable communications among PS users and officers over LTE. Firstly, MCPPT was introduced in 3GPP Release 13 to provide only voice services. In latency perspective, the mouth-to-ear latency (time gap between spoken and listened by a user ) should be less than 300 ms, whereas allocated bandwidth for each country is discussed in Table \ref{Spectrum For PS}. MCPTT in Release 13 includes features such as user authentication and service authorization, group calls on-network and off-network (broadcast group calls, emergency group calls, emergency alerts). 3GPP Release 14 brings further enhancements by introducing MC video and data services, short data service (SDS) and file distribution (FD) on-network services. Currently, Release 15 is in progress and is expected to be finalized by the end of 2018. Some of the working items in Release 15 are: enhanced MCPTT group call setup procedure with MBMS, and inter-working with TETRA and other legacy systems.
Hence, for MCPTT numerous future research challenges exist, which need researchers' attention.

\subsection{Isolated E-UTRAN operation for PS (IOPS) LTE}\label{Isolated E-UTRAN operation for public safety (IOPS)}
The resilience of PS networks can be enhanced by enabling LTE-based isolated E-UTRAN operation for PS (IOPS). This concept is introduced to continue communications even when the backhaul connectivity to the core network is lost \cite{oueis2017overview}. It comprises of isolated E-UTRAN, local evolved packet core (EPC), backhaul links, and an application server as shown in Fig. \ref{PS Architecture}(d). The isolated E-UTRAN has a nomadic eNB (NeNB) which has the capability to move and provide communication links during the emergency and disaster situations. IOPS provides resilience and service availability for networks during a disaster situation. The isolated-EUTRAN is either an E-UTRAN without normal connectivity with EPC or a NeNBs with EUTRAN functionality \cite{access3gppIoEUTRAN}. The NeNB is intended for PS-LTE to provide coverage extension and increase capacity. IOPS can provide services even if the backhaul connection to the centralized macro core is completely or partially lost. IOPS provides mission critical communication (MCC) service to PS users via isolated base stations without backhaul communications. In \cite{access3gppIoEUTRAN}, 3GPP summarizes the various IOPS scenarios with no, limited, signaling only, and normal backhaul. 

\subsection{Public warning system (PWS) in PS-LTE}\label{Public Warning System}
The public-warning system (PWS) as shown in Fig. \ref{PS Architecture}(e) is one of the important use case for PS-LTE. For example, earthquake sensor nodes are installed to gather the shock information and transmit it to the PS officers to enable PS operations. The PS officers request mobile operators to broadcast public warning alerts to the users in their vicinity. Public warning system (PWS) is an alert based system which is used for the delivery of short messages in case of emergency or disaster situations for the purpose of PS. This notification alert can be used for many purposes such as recovery of missing persons, things, documents, providing information about the shelter position in a tsunami or other man-made disaster. Many PWSs have been deployed all over the world such as commercial mobile alert system (CMAS), earthquake and tsunami warning system (ETWS), Korean public alert system (KPAS), European (EU)-ALERT, and Austria public alert system. The notification latency varies with these systems, but usually it should be delivered within 4 seconds to the user in the notification area.

\begin{figure*}[!hb]
 \centering
  \begin{center}
    \includegraphics[width=10cm,height=9.5cm,angle=0]{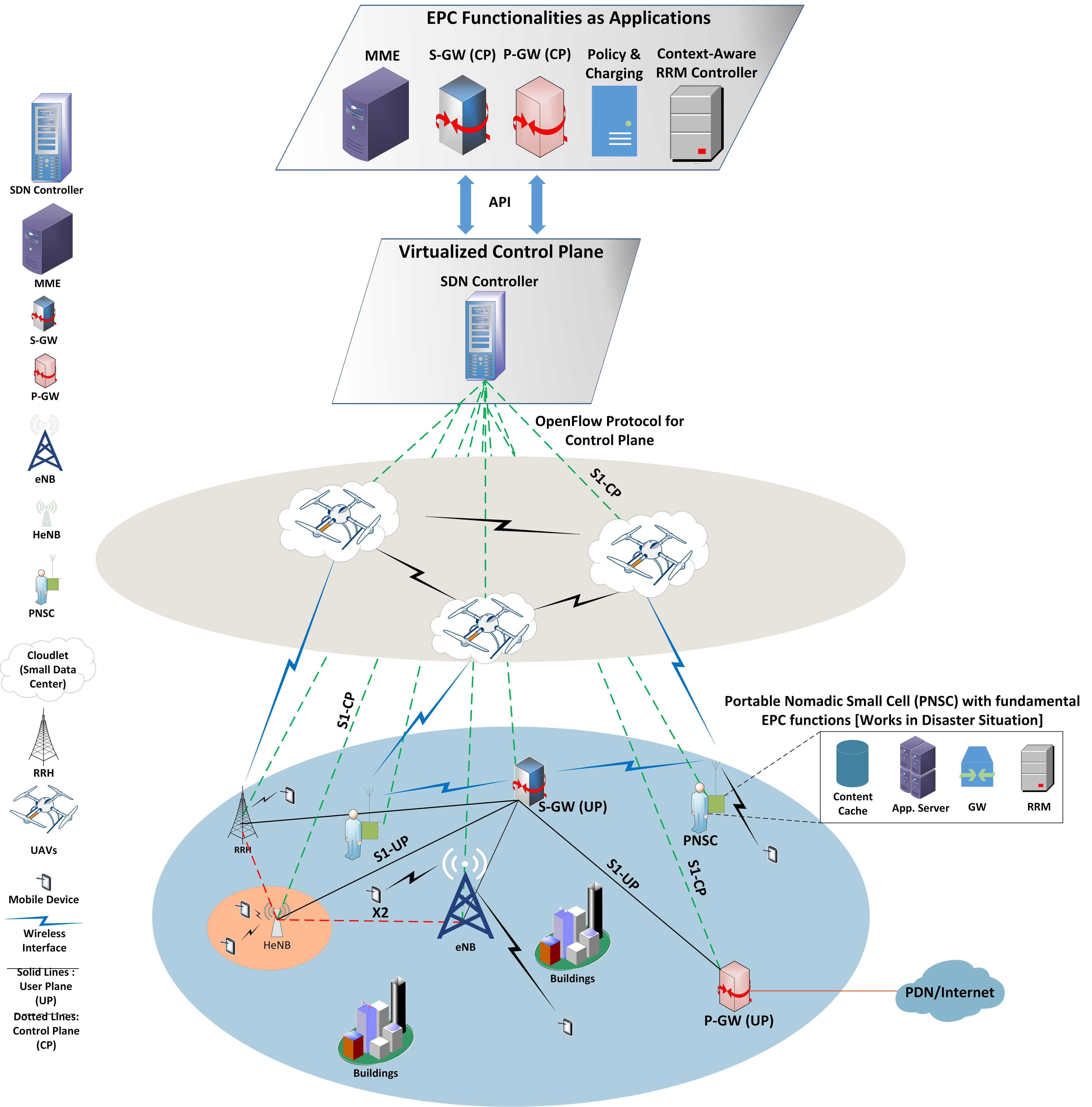}
  \end{center}
  \caption{Proposed disaster resilient architecture for PS-LTE communications.}
  \label{Architecture}
\end{figure*}

\subsection{Priority services in PS-LTE}\label{Priority Services}
The provisioning of priority services to PS users is necessary in disaster situation. For instance, unlike the conventional mobile users during a disaster situation, we need to prioritize the PS users to enable MCC services as shown in Fig. \ref{PS Architecture}(f). During the disaster situation, the MCC services are related to police car, ambulance, and fire brigade that need prioritization for timely provisioning of emergency services. Prioritization of emergency users, compared with non-PS users, in terms of radio resource allocation and connection establishment in PS-LTE is required to efficiently tackle an emergency and disaster situation. 

\section{Proposed Disaster Resilient Architecture for Delay Sensitive Public Safety Communications}\label{architecture}
3GPP presents the internet protocol (IP)-based EPS architecture to support packet-switched services. The EPS network architecture consists of E-UTRAN and EPC. 
E-UTRAN is the radio access part which connects users with EPC, where EPC controls the establishment of the bearer services and other important core network operations. 
The network elements in EPS are connected by standardized interfaces to provide inter-operability among vendors. The EPC consists of logical nodes: P-GW, S-GW, MME, HSS, and PCC. 
P-GW allocates IP addresses and maintains QoS provisioning. S-GW passes on IP packets and also terminates the interface towards E-UTRAN, manages eNB handovers, and cares about mobility interface. MME is the control node for LTE radio access that has responsibilities like user tracking and P-GW, S-GW selection. PCC manages policy and charging enforcement function (PCEF). X2 interface is used to connect eNBs with each other, whereas S1 interface is used to connect eNBs with EPC. 

The existing LTE architecture discussed above has limitations of implementation in PS-LTE. The two major limitations are: no central management of network devices and high latency because processing needs to be done from EPC. For instance, the two important use cases for PS-LTE such as lifeline communications and ultra-reliable communications cannot be implemented under the existing LTE architecture. The lifeline communications and ultra-reliable communications demand energy efficiency and low-latency, respectively. For example, for ultra-reliable communications, the pubic safety organizations demand real time video and high quality pictures transmission with very low-latency. Current LTE architecture is also not appropriate because of low flexibility, high delays, and no central control. 

To overcome these limitations, we propose a three-layered disaster-resilient PS LTE (DR-PSLTE) architecture that consists of 1) The \textbf{SDN-layer} centrally manages the network synchronization and resource management of the disaster affected area \cite{7533456}. Moreover, it also centrally manages the control signals to the lifeline provisioning PS officers. The SDN layer has EPC components implemented as application such as radio resource management (RRM) entity, control plane S/P-GW, and PCC functions. The deployment of cloud-based services and other services at the network edge results in huge complexity, this can be resolved by a central control mechanism which has the capacity to orchestrate the distributed environment. The central SDN controller, which is transparent to the end-user has the intrinsic capability to align with all these requirements and to mitigate the barriers that prevent edge computing to reach its full potential. In SDN architecture, the southbound interface is the OpenFlow protocol. The main function of this interface is to enable communication among SDN controller and various network nodes like UAV, eNodeB, and other small cells, so that they can discover the network topology, define network flows, and implement requests send to them from northbound application program interface (API). The northbound API is used to for communication between the SDN controller and the application or other higher layers as depicted in in Fig. \ref{Architecture}. Contrary to existing architecture, proposed SDN-based architecture has replaced S1 (between  MME and eNB), S11 (between MME and  S-GW) and S5 (between S-GW control and P-GW control) by the OpenFlow protocol. Whereas, the interfaces S1 (from  eNB  to  S-GW data) and  S5 (from  S-GW data  to  P-GW data)  are controlled  by  the  existing  3GPP  protocol  of  the  LTE  architecture.  The proposed architecture does not demand any change in the radio hardware at the base station. 

2) \textbf{UAV cloudlet layer} is comprised of UAV ad-hoc network to enable emergency communication links for PS users. Moreover, the UAVs are equipped with cloudlet (small data centers) to provide the EC services. It brings most of the UP and CP processing closer to the edge (end user), which results in low latency compared to the centralized only data processing. To empower edge computing capability at UAVs, they are mounted with cloudlet which can provide application offloading opportunities to users (security personal, mobile users) in the disaster-hit areas. Hence, UAVs can enable edge computing even in the absence of a wireless communication infrastructure. UAVs accompanied with cloudlet have many applications and have an imperative role in emergency relief and disaster response. For instance, users' (security personal) with limited processing capabilities devices can benefit from the cloudlet assisted implementation of data analytics application for the assessment of the status of victims using high definition image or video, hazardous terrain and structures, offload user's computationally heavy tasks that in turn reduce the battery consumption of their hand held devices. The UAV accompanied by cloudlet enable distributed processing to reduce the end-to-end latency in the key enabling emergency services such as MCPPT, PWS, and ProSe in PS-LTE.  This layer can also provide emergency backhaul communication links, when the existing backhaul is not responding and eNBs are in IOPS mode. Other benefits include the efficient management of local radio resources, ease in network up-gradation, and low operating costs.

3) \textbf{Radio access network (RAN) layer} consists of various types of eNBs such as home eNBs (HeNB), remote radio head (RRH), portable nomadic small cell (PNSC), and the user-plane gateways, and provide the radio services to the emergency users. Hence, DR-PSLTE architecture brings UP and parts of the CP functionality to the edge servers, whereas some centralized controllers are still managed by SDN to have the global view of EPC and other PS-LTE enabling functionality.

\begin{figure*}[!h]
\centering
\subfloat[Architectures delay comparison]{\includegraphics[width=7cm,height=6cm,angle=0]{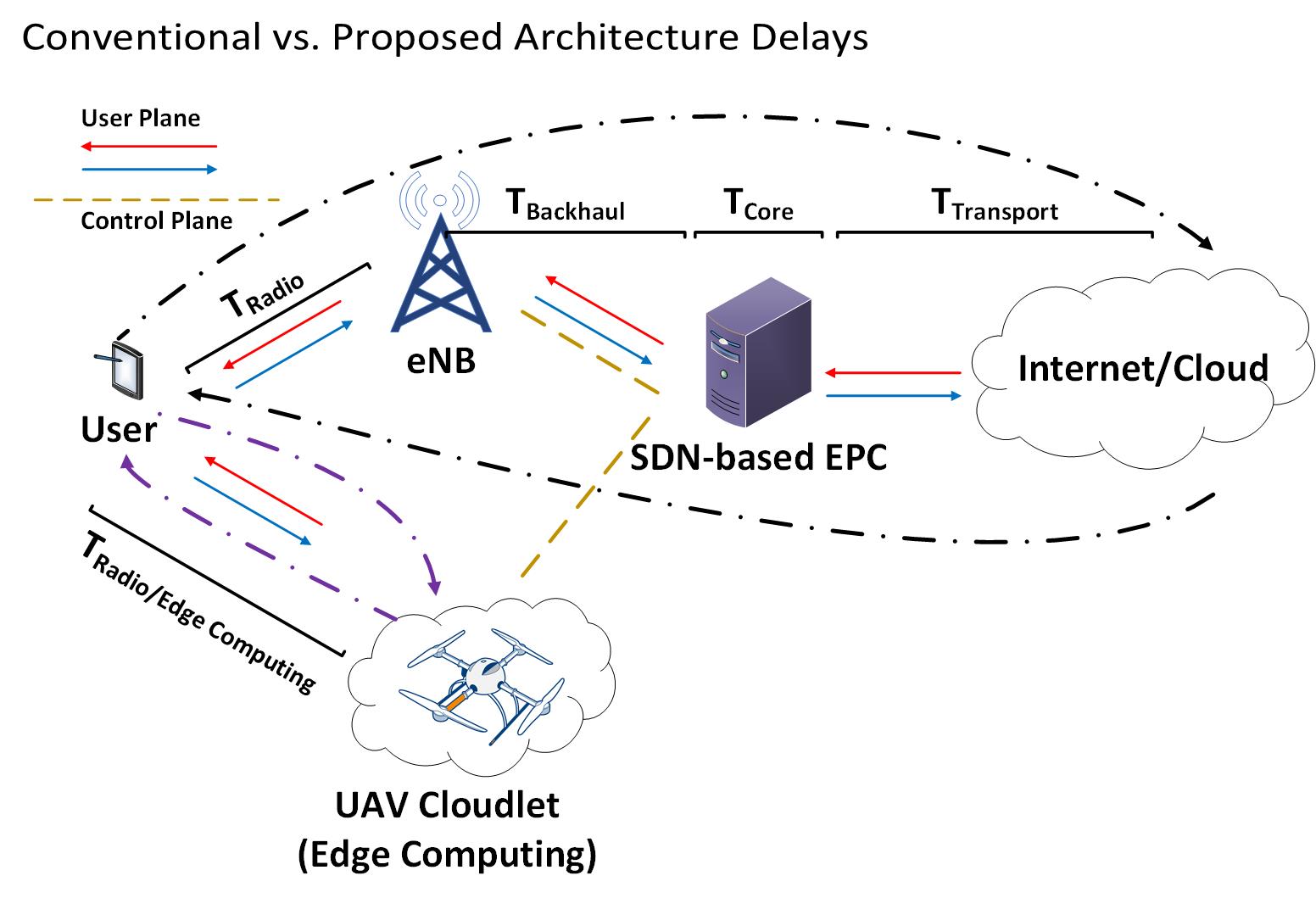}}%
\hfil
\subfloat[Computation cycles vs. system delay]{\includegraphics[width=7cm,height=5cm,angle=0]{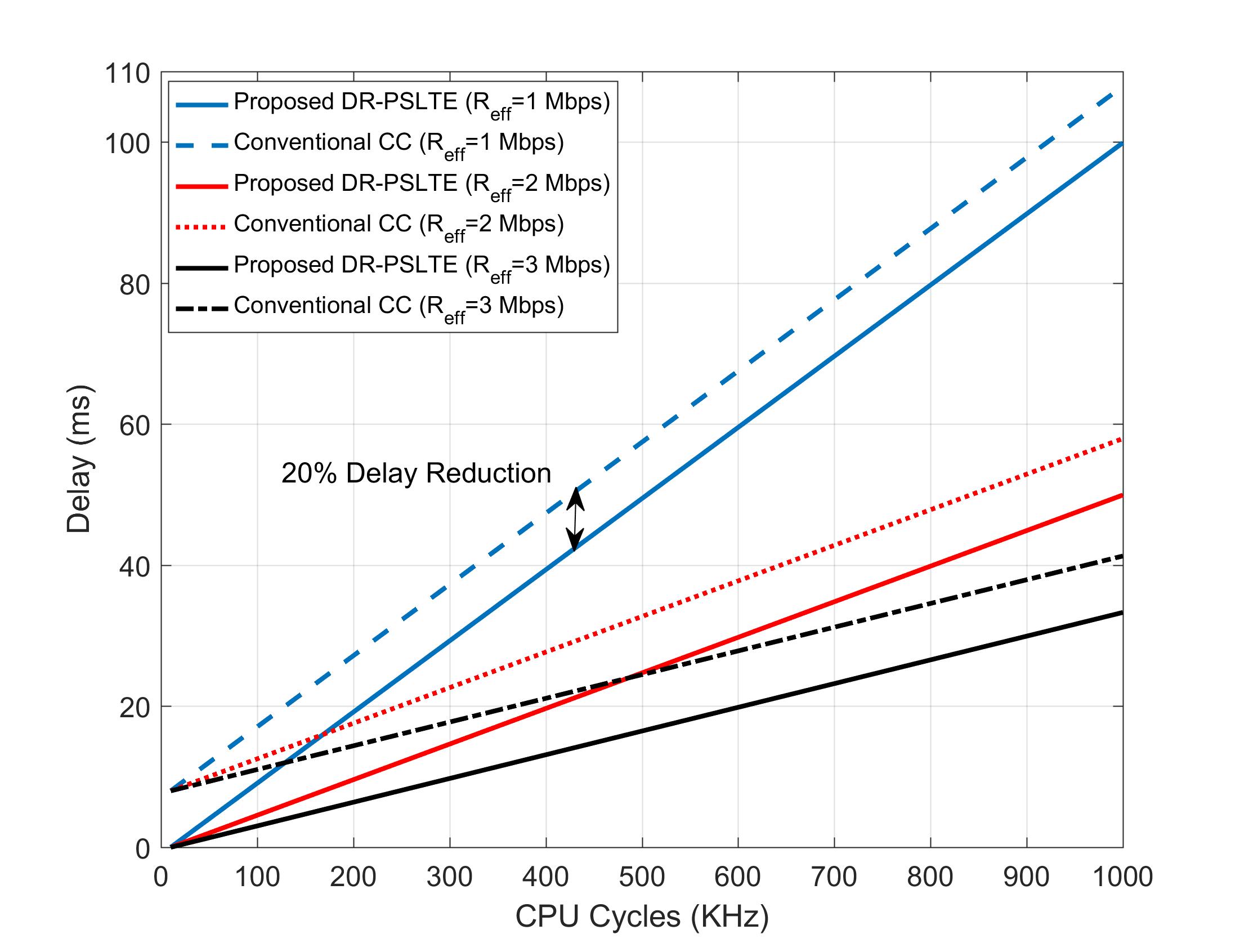}}%
\hfil
\subfloat[Transmitted data vs. energy consumption]{\includegraphics[width=7cm,height=5cm,angle=0]{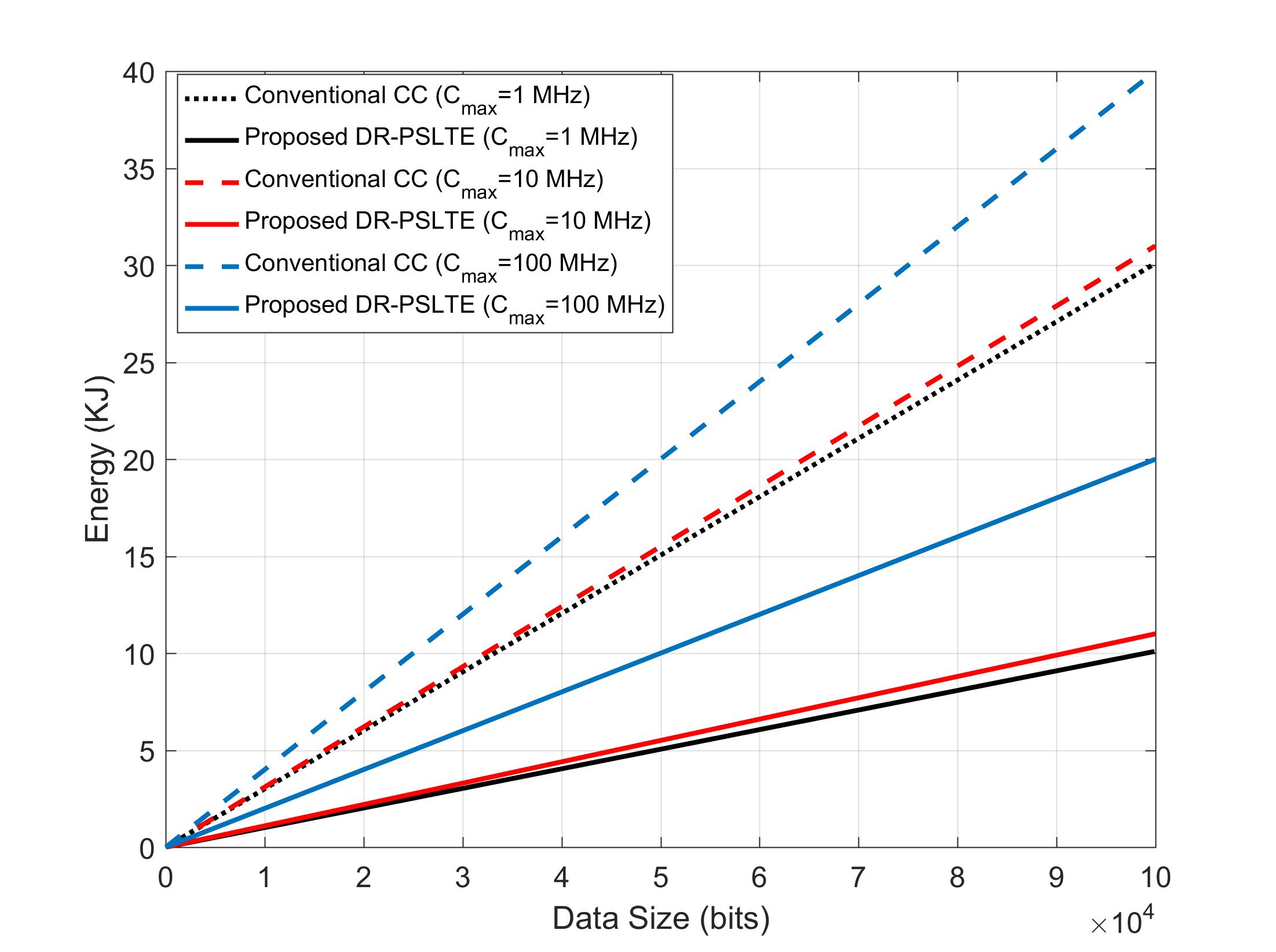}}%
\hfil
\subfloat[Effect of increasing edge nodes]{\includegraphics[width=7cm,height=5cm,angle=0]{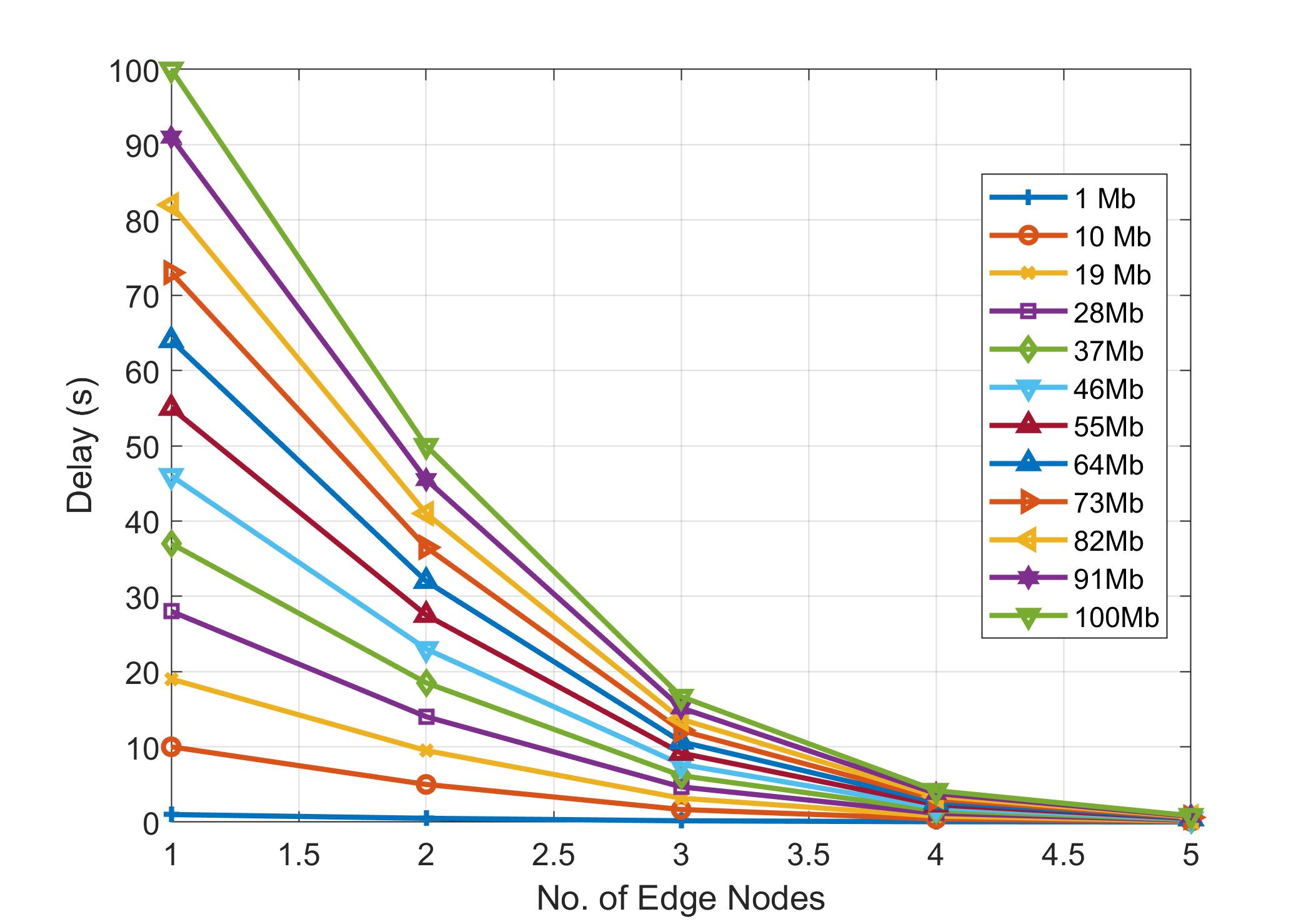}}%
\caption{Single node representation of DR-PSLTE Architecture and results: (a) Proposed vs. conventional architecture single node delay diagram; (b) No. of computation cycles vs. system delay; (c) Transmitted data vs. energy consumption;(d) Effect of increasing edge node on system delay.}
\label{delays}
\end{figure*}

\section{Performance Evaluation and Open Challenges}
In this section, we compare the performance of the proposed DR-PSLTE architecture and conventional cloud computing architecture in terms of energy consumption and latency. We also highlighted the open research challenges after the deployment of DR-PSLTE in public safety environment.

\subsection{Numerical Results}
To design a low latency network, we consider an integrated software-defined network and a UAV cloudlet-based three layered network architecture to enable centralized control and edge computing functionality, respectively as shown in Fig. \ref{delays} (a). To enable edge offloading for each user $u \in \mathbb{U}=\{1,....,U\}$, $K$ UAVs that acts as edge nodes are deployed. We assume that each $j\in \mathbb{J}=\{1,....,J\}$ computation task is defined by ($C_j,D_j$), where $C_j$ and $ D_j $ are the number of computation cycles and the data size, respectively that needs to be forwarded. Two main types of latencies are considered here: \textit{computational} and \textit{communication latency}, where these latencies depend on computation capability of a node and on link bandwidth (or achievable effective data rate $R_{eff}$), respectively. To accomplish each task, two strategies are adopted: 1) tasks offloading to the conventional CC architecture which results in large delay, 2) computing the latency sensitive tasks at UAV-enabled edge computing server.  

The CC enabled system requires the information to pass through several networks that contributes to delays such as radio network ($T_{Radio}$), backhaul network ($T_{Backhaul}$), core network ($T_{Core}$), and transport network ($T_{Transport}$) \cite{parvez2018survey}. Whereas in the proposed EC based DR-PSLTE architecture, it has only radio network ($T_{Radio/Edge Computing}$) delay besides the computation delays as shown in Fig. \ref{delays} (a). 

To prove the efficacy of the proposed proposed DR-PSLTE architecture, we compare the performance in terms of computing and communication latency, and energy consumption. 
For the proposed DR-PSLTE architecture delay and energy for a single node is computed as

\begin{align}
\label{eq1}
T^{DR-PSLTE}_j=&\left(\frac{C_j}{F_{EC}}+\frac{D_j}{R_{eff}}+ T_{Radio/Edge Computing}\right), \\ \nonumber
E^{DR-PSLTE}_j=&(C_j \times e^{cpu}_{EC}+D_j \times e^{d}_{EC}),
\end{align}
where $ F_{EC}$ denotes the CPU frequency of the EC server, $ R_{eff} $ is the achievable effective data rate, $e^{cpu}_{EC}$, and $e^{d}_{EC}$ is the energy consumed per CPU cycle and energy consumed per data unit by EC server, respectively. The important parameters used to obtain numerical results are summarized in Table \ref{table2}, where low value of $ e_{EC} $ represents that EC has no restrictions in terms of energy.

Similarly, the delay occurred and energy consumed in the conventional CC architecture is calculated as

\begin{align}
\label{eq2}
T^{CC}_j=&(\frac{C_j}{F_{CC}}+\frac{D_j}{R_{eff}}+ T_{Radio}+ T_{Backhaul}\\\nonumber +&T_{Core} + T_{Transport}), \\ \nonumber
E^{CC}_j=&(D_j \times e^{d}_{EC}+D_j \times e^{d}_{BS}+C_j \times e^{cpu}_{c}+D_j \times e^{d}_{c}),
\end{align}
where $ F_{CC}$ denotes the CPU frequency of the CC server, and $e^{cpu}_{c}$ is the energy consumed per CPU cycle, $e^{d}_{BS}$ and $e^{d}_{EC}$ are the energy consumed per data unit transmission by BS and edge computing server, respectively, and the values used for numerical results are summarized in Table \ref{table2}.

\begin{table}[!h]
\caption{Numerical parameters} 
\centering 
\begin{tabular}{c c }
\hline 
\textbf{Parameters} & \textbf{Values} \\ [0.5ex] 
\hline 
$F_{CC}$, $F_{EC} $ & 50 GHz, 5 GHz\\
Computation Cycles $C_j$ &	$[10,....,1000] \times 10^3$ \\
Data Size $D_j$ &	$[1,....,100]\times 10^3$ \\
$R_{eff}$ & \{1, 2, 3\} Mbps \\
$T_{Radio}$, $T_{Backhaul}$, $T_{Core}$, $T_{Transport}$	& 2 ms, 2 ms, 1 ms, 3 ms \cite{parvez2018survey}\\
$e^{d}_{BS}$, $e^{cpu}_{c}$, $e^{d}_{c}$ & 0.1, 0.1, 0.1\\
$e^{cpu}_{EC}$, $e^{d}_{EC}$ & 0.1, 0.1 \cite{messous2017sequential,messous2017computation}\\
\hline 
\end{tabular}
\label{table2} 
\end{table}

From Fig. \ref{delays} (b) we can notice that the proposed DR-PSLTE outperforms the conventional CC architecture by deploying UAV cloudlet to enable EC. Thus, due to shorter communication loop the processing delay is reduced by 20\%, which in turn helps to promptly process emergency computation services at the edge as compared with the conventional CC to save the victims. This trend continues even by varying the effective data rate $ R_{eff}= \{1, 2, 3\}$ Mbps or the available system bandwidth. That is, by increasing the available bandwidth or effective data rate, delay reduces when compared at the same processing frequency.

To verify how the processing delay of the proposed system is affected with the increase in the number of UAVs deployed at the edge. We tested the proposed system when the total data lies in a range of $[1, 2,...,100]$ Mb as shown in Fig.\ref{delays} (d). Simulation results show that the processing delay decreases as the number of deployed UAV at the edge increases. We can notice that the decrement in delay is very small when data to be transmitted is small (for instance 1 Mb) or equal to the link effective data rate $R_{eff}=1$ Mbps. Thus, single UAV deployed at the edge can easily process the data within the deadline as no queue will be created. However, for large amount of data (such as from 10-100 Mb), this delay increases significantly for using single edge UAV as all data needs to be processed on a single edge node, which results in large queues. Hence, deployment of more number edge UAVs results in significant reduction of processing delay as task will be distributed on various edge nodes. This will result in little increase in communication delay but that delay would be constant for all nodes because of stable communication link, and would not contribute much in the system delay as depicted in Fig.\ref{delays} (d).

We also compared the proposed DR-PSLTE architecture energy consumption performance with the conventional CC architecture. Fig. \ref{delays} (c) demonstrate that energy consumption increases when more data needs to be transmitted. However, we can notice that the proposed DR-PSLTE energy consumption is significantly less than conventional CC when compared at various range of maximum computation cycles, that is $C^{max}_{j}=\{1, 10, 100\}$ MHz because of short communication latency. Hence, this system can be adopted in the PS-LTE system to enable emergency communications. These simulations are performed by ignoring the UAV battery constraints and network management issues.

\subsection{Open Research Challenges}
Despite the tremendous advantages brought by the proposed DR-PSLTE architecture by enabling low-latency and low-energy emergency communications, there are still several open challenges. 

\textbf{Integration of various communication technologies}: The integration of various communication technologies in layered architecture poses inter-working challenges as each technology works on different communication protocols. Therefore, the data exchange among these nodes has to be conducted in multi-protocol environment. So, research in designing protocols that can efficiently work in this environment is the challenge of utmost importance.

\textbf{Channel modeling}: There is lot of dynamics involved in communication among the ground user and UAV deployed for edge computing. The main reason is the highly fluctuating wireless channels among ground user to UAV and UAV-UAV. To solve these challenges, stochastic channel modeling can be considered a potential solution where various 3D channel parameters are fine tuned for the fluctuating environment.

\textbf{UAV placement and Resource allocation optimization}: The UAV cloudlet placement at the edge needs is a important challenge that needs proper investigation. Therefore, 3D placement of UAV at the edge with minimum processing time and least energy consumption constraints needs attention to obtain high edge computing efficiency. Since UAV has an important role in DR-PSLTE as it performs various computing task at the edge. Thus, resource allocation, interfacing with other devices, management of important control functions, and task scheduling needs to be jointly optimized to obtain the desired performance. Moreover, the UAV trajectory and user mobility must be focused to obtain the efficient results. Since, UAV are sensitive to the energy constraint, while designing scheduling algorithms this constraint should be incorporated to guarantee the network performance. Furthermore, the management and synchronization of huge number of deployed UAVs needs intensive research. In this article, we highlighted most important challenges and research directions that needs proper attention and extensive research works.

\section{Conclusion}\label{Concluding Remarks}
3GPP LTE is a key enabler for the emergency communication services in PS situations. In this article, we briefly discussed the communication services enabler in PS-LTE. Moreover, the 3GPP status of various PS-LTE related services such as proximity services, emergency call, IOPS, public warning system, and mission critical services were presented. Finally, we proposed a disaster-resilient architecture for PS-LTE that plays a key role in providing the emergency communication services in disaster affected areas. It combines the benefits of software-defined networks and UAV cloudlets, which help to meet the energy and latency requirements of PS users by enabling centralized and distributed processing. Simulation results shows that the proposed DR-PSLTE architecture achieved 20\% less delay and has low energy consumption as compared to the conventional centralized computing. In the future, we will extend our work to address the limitations of UAV placement and group management for optimum system performance. 
\bibliography{References}\label{psbib.bib}
\bibliographystyle{IEEEtran}{}

\begin{IEEEbiography}[{\includegraphics[width=1in,height=1.25in,clip,keepaspectratio]{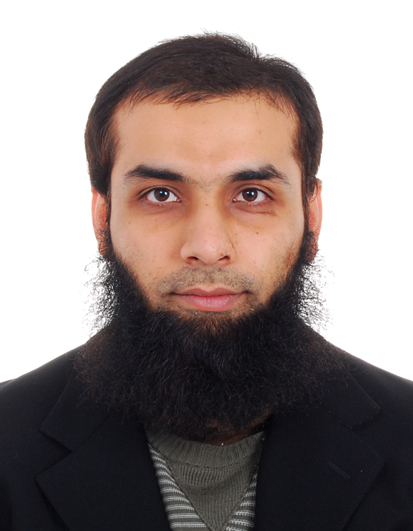}}]{Zeeshan Kaleem}
(zeeshankaleem@gmail.com) is an assistant professor in the
Department of Electrical Engineering, COMSATS Institute of information Technology,
Pakistan. He received his Ph.D. degree in electronics engineering from
INHA University, Korea, in 2016. He has authored several peer-reviewed journal/
conference papers and holds 18 U.S./Korea patents. He is an Associate Editor of
IEEE Access and IEEE Communications Magazine. His research interests include
device-to-device (D2D) communications/discovery, unmanned air vehicles (UAV),
and resource allocation in fifth-generation (5G) networks

\end{IEEEbiography}

\begin{IEEEbiography}[{\includegraphics[width=1in,height=1.25in,clip,keepaspectratio]{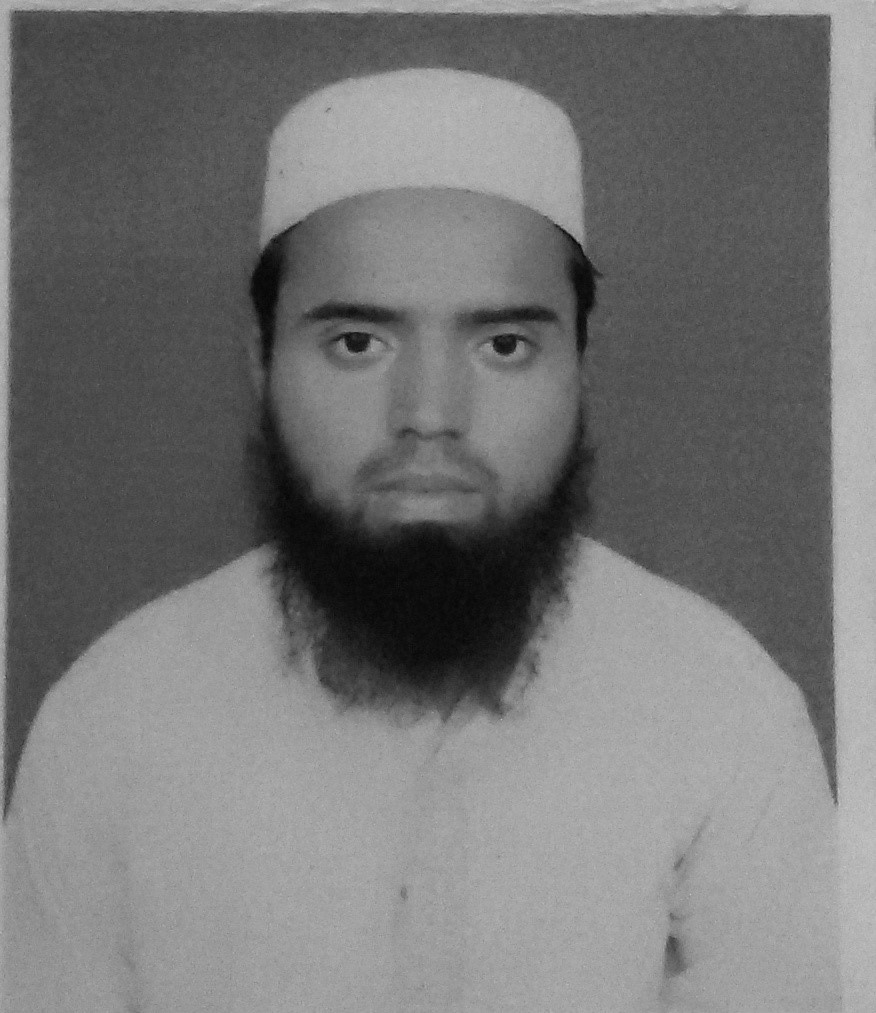}}]{Muhammad Yousaf}
received his B.Sc Degree in Electronics Engineering from the University of Engineering and Technology (UET) Taxila, Pakistan, in 2014. He received M.Sc Degree in Electrical Engineering from the COMSATS Institute of  Information Technology (CIIT) Wah, in 2017. Currently, he is a PhD student in Telecommunications Department in UET Taxila. His current research interests include wireless communication, image processing, and PS long-term evolution (PS-LTE).
\end{IEEEbiography}

\begin{IEEEbiography}[{\includegraphics[width=1in,height=1.25in,clip,keepaspectratio]{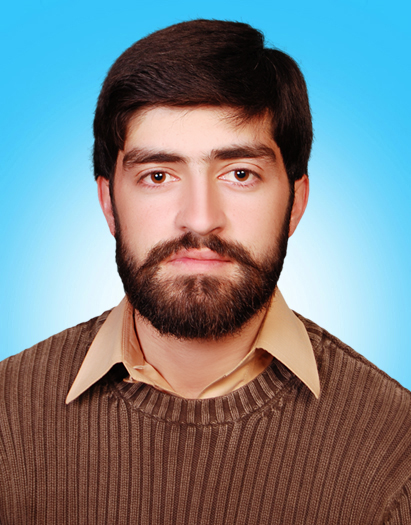}}]{Aamir Qamar}
received his bachelor degree in Electronics Engineering from NWFP University of Engineering and Technology (UET), Pakistan. In 2012 he received MS degree in Electrical Engineering with Specialization in Signal Processing and Wave Propagation from Linnaeus University, Sweden. In 2016 he was awarded PhD degree in
Electrical Engineering from Chongqing University, China. During his PhD he was with the State and Key Laboratory of Power Transmission $ \& $System Security and New Technology, Chongqing University. His research interests are in the field of Signal Processing and Electromagnetic. He is currently working on the performance
analysis of substation grounding grid and public-safety communications.

\end{IEEEbiography}

\begin{IEEEbiography}[{\includegraphics[width=1in,height=1.25in,clip,keepaspectratio]{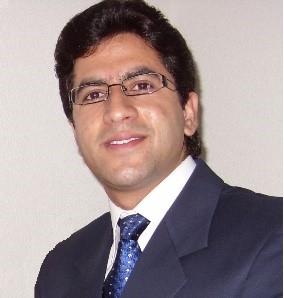}}]{Ayaz Ahmad}
\textbf{(S' 08, M' 15, SM' 16)} is Assistant Professor in COMSATS Institute of Information Technology, Wah Cantt., Pakistan. He received his Master and Ph.D degrees both in wireless communication form SUPELEC, Gif-sur-Yvette, France, in 2008 and 2011, respectively. He has authored/co-authored several scientific publications. He has also edited a book and has served as guest editor for two special issues in IEEE Access. He is Associate Editor with IEEE Access and Springer Human-centric Computing and Information Sciences.

\end{IEEEbiography}

\begin{IEEEbiography}[{\includegraphics[width=1in,height=1.25in,clip,keepaspectratio]{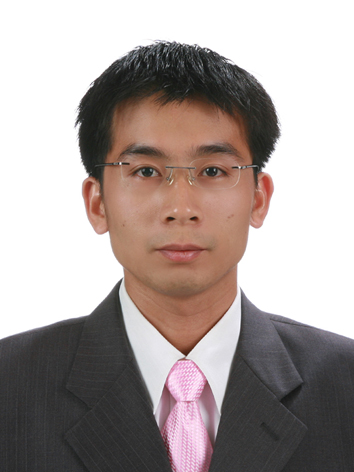}}]{Trung Q. Duong}
\textbf{(S'05 M'12 SM'13)} is an Assistant Professor at Queen's University Belfast, U.K. He has authored 110 IEEE journals articles. He is currently serving as an Editor for IEEE Trans on Wireless Communications, IEEE Trans on Communications. He was awarded the Best Paper Award at IEEE VTC 2013, ICC 2014, IEEE GLOBECOM 2016, IEEE DSP 2017. He is a recipient of prestigious Royal Academy of Engineering Research Fellowship from 2016 to 2021.

\end{IEEEbiography}

\begin{IEEEbiography}[{\includegraphics[width=1in,height=1.25in,clip,keepaspectratio]{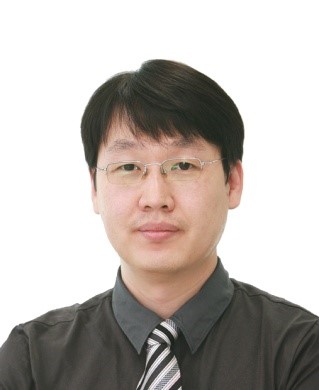}}]{Wan Choi}
is Professor of School of Electrical Engineering, KAIST, Korea. He received IEEE VT Society Jack Neubauer Memorial Award, Dan Noble Fellowship Award, and IEEE Communication Society AP Young Researcher Award. He is an Executive Editor for IEEE Trans. Wireless Communications, and Editor for IEEE Trans. Vehicular Technology. He served as Editor for IEEE Trans. Wireless Communications and IEEE Wireless Communications Letters, and Guest Editor for IEEE Journal on Selected Areas in Communications.

\end{IEEEbiography}

\begin{IEEEbiography}[{\includegraphics[width=1in,height=1.25in,clip,keepaspectratio]{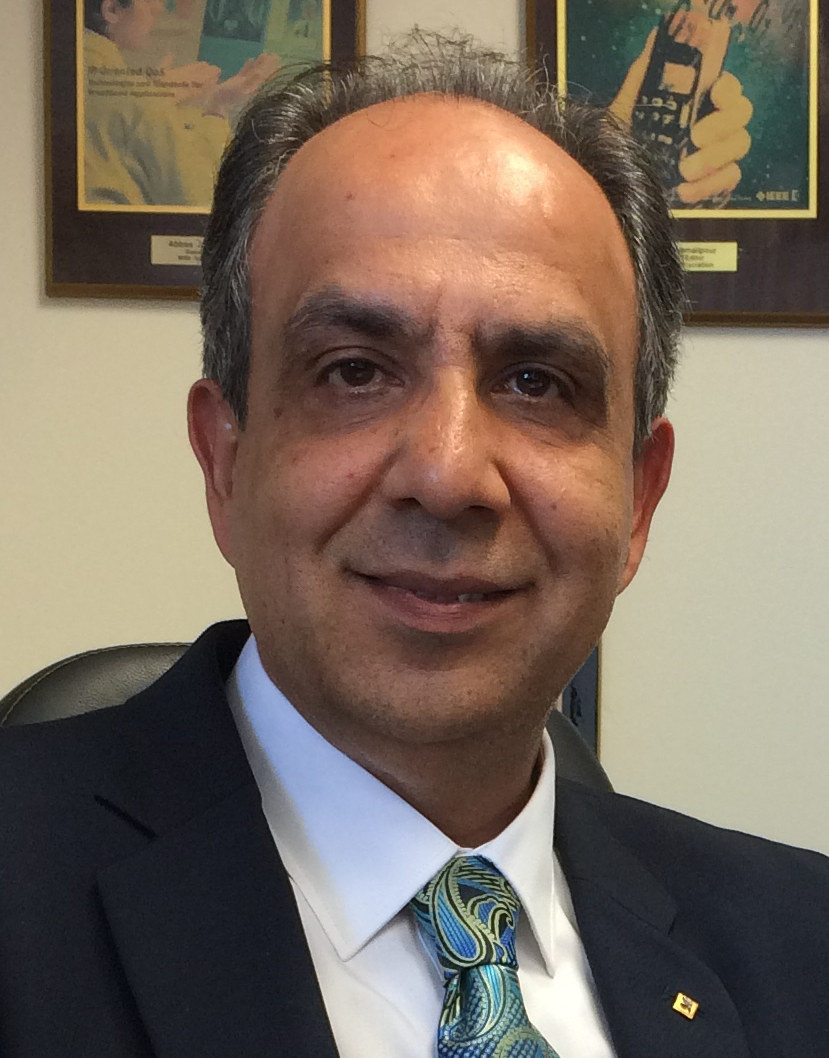}}]{Abbas Jamalipour}
Abbas Jamalipour (S'86-M'91-SM'00-F'07) is the Professor of Ubiquitous Mobile Networking at the University of Sydney, Australia, and holds a PhD in Electrical Engineering from Nagoya University, Japan. He is a Fellow of the Institute of Electrical, Information, and Communication Engineers (IEICE) and the Institution of Engineers Australia, an ACM Professional Member, and an IEEE Distinguished Lecturer. He has authored six technical books, eleven book chapters, over 450 technical papers, and five patents, all in the area of wireless communications. He was the Editor-in-Chief IEEE Wireless Communications (2006-08), Vice President-Conferences (2012-13) and a member of Board of Governors of the IEEE Communications Society, and has been an editor for several journals including IEEE Trans. Vehicular Technology. He has organized 32 special issues on hot topics in the field mainly in IEEE journal and magazine. He has held positions of the Chair of the Communication Switching and Routing and the Satellite and Space Communications Technical Committees and Vice Director of the Asia Pacific Board, in ComSoc. He was a General Chair or Technical Program Chair for a number of conferences, including IEEE ICC, GLOBECOM, WCNC and PIMRC. Dr. Jamalipour is an elected member of the Board of Governors (2014-16 and 2017-19) and the Editor-in-Chief of the Mobile World, IEEE Vehicular Technology Society. He is the recipient of a number of prestigious awards such as the 2010 IEEE ComSoc Harold Sobol Award, the 2006 IEEE ComSoc Distinguished Contribution to Satellite Communications Award, the 2006 IEEE ComSoc Best Tutorial Paper Award, and Distinguished Technical Achievement Award to Communications Switching and Routing, IEEE Communications Society, 2016.

\end{IEEEbiography}

\end{document}